\documentclass[twocolumn]{aastex631}
\received{\today}

\graphicspath{{figures/}}
\usepackage{aas_macros}
\usepackage{physics}
\usepackage{booktabs}
\usepackage{multirow}
\usepackage{xspace}
\usepackage{natbib}
\usepackage{fontawesome5}
\usepackage{xcolor}
\usepackage{wrapfig}
\usepackage[figuresright]{rotating} 
\usepackage[version=4]{mhchem}
\usepackage[hang,flushmargin]{footmisc} 
\usepackage[mathlines]{lineno}

\newif\ifshowrevisions

\ifshowrevisions
  \newcommand{\rev}[1]{{\bfseries #1}}
\else
  \newcommand{\rev}[1]{#1}
\fi

\makeatletter
\newenvironment{fotoexample}[2]{%
  \par\medskip\noindent\begin{minipage}{\linewidth}%
  \textbf{\href{https://arxiv.org/abs/#1}{arXiv:#1}, fig.~#2}\par\smallskip
}{%
  \end{minipage}\par\medskip
}
\newcommand{\extitle}[1]{\noindent\textit{Title:} #1\par\smallskip}
\newcommand{\excaption}[1]{\noindent\textit{Caption:} #1\par\smallskip}

\newlength{\exqueryw}\newsavebox{\exbox}
\newcommand{\qrow}[3]{#1 & \parbox[t]{\exqueryw}{\strut #2\strut} & #3 \\}
\newenvironment{querytable}{%
  \sbox{\exbox}{\begin{tabular}{@{}l l r@{}}
    Register & & Rank \\ terse & & \\ casual & & \\ vague & & \\
    detailed & & \\ notation & & \\ \end{tabular}}%
  \setlength{\exqueryw}{\dimexpr\linewidth-\wd\exbox\relax}%
  \noindent\begin{tabular}{@{}l l r@{}}
  \hline
  Register & Query & Rank \\
  \hline
}{%
  \hline
  \end{tabular}%
}

\newcommand{\checknextarg}{\@ifnextchar\bgroup{\gobblenextarg}{}}
\newcommand{\gobblenextarg}[1]
{\,\mathrm{#1}\@ifnextchar\bgroup{\gobblenextarg}{}}
\makeatother

\begin{document}

\title{\Large FOTO:  Semantic retrieval of scientific figures for Astronomy}

\newcommand{\UW}{Department of Astronomy, University of Washington, Seattle, WA, 98195}

\newcommand{\CCA}{Center for Computational Astrophysics, Flatiron Institute, 162 5th Avenue, New York, NY 10010, USA}

\author{Hurum Maksora Tohfa}
\affiliation{\UW}
\author{Francisco Villaescusa-Navarro}
\affiliation{\CCA}

\begin{abstract}
Astronomy papers communicate much of their content visually through figures, and existing literature search tools index text but not figures, leaving no way to search the literature at the level of individual plots. We present FOTO (Figure frOm Text \& illustratiOns), an open source tool that retrieves published figures from a text description. FOTO indexes figures by their captions. Each caption is concatenated with its paper title and then is embedded with a locally hosted model (bge-base-en-v1.5). Candidate figures are retrieved by dense similarity to the query, and \rev{a language model} then verifies which of the retrieved candidates actually match the request, keeping the tool honest about only returning figures that exist in the corpus. We build a caption index from the astro-ph slice of ArXivCap, covering 53,839 papers and 482,750 figure captions through June 2023, and evaluate it with a benchmark of queries written in five registers, from terse jargon to vague conversational description. \rev{FOTO outperforms the paper-level search tools Pathfinder and Semantic Scholar by a wide margin in every query register, even though the baselines only need to find the right paper while FOTO must find the exact figure. Because the benchmark queries are generated from figure captions and each tool searches its own index, this comparison shows what each tool offers a researcher today rather than comparing retrieval methods on equal terms. Retrieval is strongest for well-specified queries and weaker for vague ones, and reranking the retrieved candidates with a language model judge substantially improves the ranking, with openly available judges performing as well as a frontier model.} 
\end{abstract}
\section{Introduction} 
\label{sec:intro}

As one of the oldest scientific disciplines, astronomy has accumulated a large and rapidly growing body of literature. Repositories like the Astrophysics Data System \citep[ADS;][]{kurtz2000ads} and the arXiv provide lasting access to this record, and the rate at which it grows keeps accelerating: 18,660 papers were posted to astro-ph in 2025, up from 16,333 the year before, corresponding to roughly 70 new papers per working day \citep{2026arXiv260212303L}. With new facilities like JWST, Rubin, and the gravitational wave detectors driving output across every subfield, keeping pace with the literature has become difficult even for experienced researchers, and harder still for those whose work spans multiple subfields. A large fraction of this information is communicated visually. The average astro-ph paper contains 9 figures \citep{2026arXiv260212303L}, which amounts to more than 150,000 new figures entering the literature each year, and these figures often summarize the central results of their papers: parameter constraints, spectra, light curves, correlation functions, and scaling relations. Existing search services in ADS and arXiv support keyword and metadata queries over titles, abstracts, and full text, and Semantic Scholar \citep{2023arXiv230110140K} adds relevance signals from the citation graph. Semantic Scholar also pulls figures out of papers so they can be browsed one paper at a time, but none of these services lets a user search for a figure by describing what it shows. Keyword search is poorly suited to the task since a researcher describing a plot rarely reproduces the vocabulary of its caption, which is dense with paper specific notation, named surveys, and jargon, so the query and the target share few exact terms for a syntactic match. A large language model can get past this vocabulary gap, since a user can describe a plot in plain words and let the model work out what they mean. On its own, though, an LLM is a shaky way to find a specific published figure as when asked for a plot, it can point to one that does not exist or describe a figure it never actually saw \citep{huang2023hallucination}, and searching a whole corpus by sending every candidate through a large model is slow and costly. In research this kind of error matters, since a figure that was never published is worse than no answer at all, so what we want is a way to find real figures that stays tied to the published record rather than invented by the model.

This problem is not specific to astronomy, but astronomy is well positioned to address it. The field maintains a largely open and centrally indexed literature, which has already supported a series of machine learning efforts on the text side. Domain adapted language models such as astroBERT \citep{grezes2021astrobert} and AstroLLaMA \citep{nguyen2023astrollama} demonstrated that models trained on astronomical corpora handle the jargon and named entities of the field, and \texttt{Pathfinder} \citep{iyer2024pathfinder} provides semantic search over \rev{385,166} peer reviewed abstracts by embedding queries and papers in a shared vector space, retrieving literature by meaning rather than keyword overlap. In parallel, general purpose text embedding models \citep{2019arXiv190810084R, xiao2023cpack} have become small enough to run locally while remaining competitive with proprietary APIs, and vision language models are now good enough at reading scientific plots to check whether a candidate figure matches a short text description \citep{2021arXiv210300020R, roberts2024scifibench}. Recent work has also started to study figure retrieval directly. SciMMIR \citep{wu2024scimmir} builds a large set of figure and caption pairs from across the sciences and retrieves figures by embedding the figure image and matching it against text. That work treats the problem as image to text matching, and it reports that off the shelf models need fine tuning on in domain data before they do well. We take a caption first route instead and focus on astronomy. \rev{Figure search tools also exist outside this line of research. In biomedicine, Open-i \citep{demnerfushman2012openi} indexes figures from the open-access literature by their captions and the passages that mention them, together with image features, and accepts both text and image queries. In astronomy, the Astronomy Image Explorer \citep{aie2013} offers keyword search over figures published in the AAS journals, matching query terms against captions and weighting the article title, abstract, and the passages that refer to each figure. Unlike the Astronomy Image Explorer, FOTO retrieves figures by semantic similarity rather than keyword overlap and indexes peer-reviewed astro-ph papers across journals, and unlike both tools it verifies candidates with a language model, which supports the descriptive and vague queries researchers use when looking for a figure they only partly remember.}

We present \texttt{FOTO} (Figure frOm Text \& illustratiOns), an open source tool that retrieves published figures from a text description. The pipeline has three stages. First, figure captions, concatenated with their paper titles, are embedded with a locally hosted open model \citep[bge-base-en-v1.5;][]{xiao2023cpack}. Candidate figures are then retrieved by dense similarity to the query. Finally, \rev{a language model} inspects the surviving candidates and scores whether each one actually matches the request. This split is what keeps the results honest: every figure returned is a real one pulled from the index rather than something the model made up, and the \rev{verifier} only has to check a short list of candidates rather than reason over the whole corpus. \rev{Users can also supply a rough sketch, which a vision-capable verifier reads at this stage.} Because the embedding model runs locally, a user needs only a single API key for the verification model, and the retrieval index can be rebuilt from public data.

To evaluate the system we build a caption index from the astro-ph slice of the \rev{ArXivCap} dataset \citep{li2024multimodalarxiv}, covering 53,839 papers and 482,750 figure captions through June 2023. Real queries do not look like captions, so we construct a multi register benchmark in which each target figure is queried in five styles, ranging from terse jargon to vague conversational descriptions.  We use this benchmark to validate the main design choices such as what text to embed, whether rewriting captions with an LLM helps, which embedding model to use, and how performance carries over from a pilot sample to the full corpus. Embedding the title of the paper the figure comes from together with the raw caption dominates every alternative we tried, caption rewriting adds no measurable gain, and a small local embedding model outperforms a paid API. At full corpus scale retrieval remains strong for specific queries (recall at 20 of 0.80 for detailed and terse queries), while vague queries are the clear failure mode (0.40 at 20, rising to 0.47 at 50). Reranking the retrieved candidates with a language model recovers much of this gap, lifting recall at 1 on vague queries from 0.07 to 0.57 when the target is among the top 50 candidates. These vague queries are the natural target for future fine tuning.

This paper is organized as follows. Section~\ref{sec:data} describes the corpus and its preparation. Section~\ref{sec:methods} presents the retrieval architecture and the verification stage. Section~\ref{sec:benchmark} introduces the multi register query benchmark and presents the results, \rev{Section~\ref{sec:availability} describes data and code availability, and Section~\ref{sec:conclusion} summarizes our work.}

\section{Data}
\label{sec:data}

Our figure corpus is built on ArXivCap \citep{li2024multimodalarxiv}, a large figure-caption dataset drawn from arXiv papers across many fields\footnote{Licensing: ArXivCap is released under CC BY-NC-SA 4.0. We keep our derived index in a separate repository under the same terms.}. ArXivCap starts from the LaTeX source of papers posted before June 2023 and keeps only papers that Semantic Scholar \citep{2023arXiv230110140K} records as published in a journal, conference, or review venue, on the assumption that peer review sets a floor on figure and caption quality. Figures and captions are pulled straight from the LaTeX source by matching the figure and caption commands, rather than from the rendered PDF as tools like PDFFigures do \citep{clark2016pdffigures}, and the images are saved as JPEG. Because of this, the captions come out as clean text with the math kept, instead of the noisier output we get from reading a figure image. Each record is stored as a chunk that is either a single figure with its caption or a group of sub-figures with their sub-captions and one overall caption. Captions are cleaned by dropping very short ones and turning LaTeX into readable text, and images that are too thin, too small, or too large are removed. In total ArXivCap holds about 6.4M images and 3.9M captions from 572K papers across 32 fields, which makes it, at the time of writing, the largest figure-caption set collected from real papers rather than synthetic plots.

For this work, we restrict ArXivCap to papers in the astro-ph category. Because our retrieval runs over text, we pull only the text columns which contains the paper titles, captions, and identifiers, rather than the full image set using DuckDB. The astro-ph covers 53,839 astronomy papers posted through June 2023, holding 488,498 figure-caption entries. After we drop captions shorter than 20 characters we are left with 482,750 usable entries.

For each figure we build the text we embed by cleaning the caption and prepending the paper title, joined with a delimiter as ``title $\bigm|$ caption''. The clean up step is a deterministic pass that removes residual LaTeX markup from the caption, stripping citation, references and font-formatting commands and math delimiters while preserving the symbols themselves (for example `\$sigma\_8\$' becomes `sigma\_8'). This prevents leftover markup from entering the embedded text. We do not rewrite or expand the captions, since as we show in Section~\ref{sec:benchmark} that rewriting does not improve performance. We embed the same cleaned text for every index we compare so that the comparisons are fair.

Our corpus doesn't contain newer papers past June 2023 as ArXivCap ends in June 2023. The peer-review filter also leaves out papers that only ever appeared as preprints, along with any paper whose source could not be parsed.

\section{Methods}
\label{sec:methods}
\begin{figure*}
\centering
\includegraphics[width=2\columnwidth]{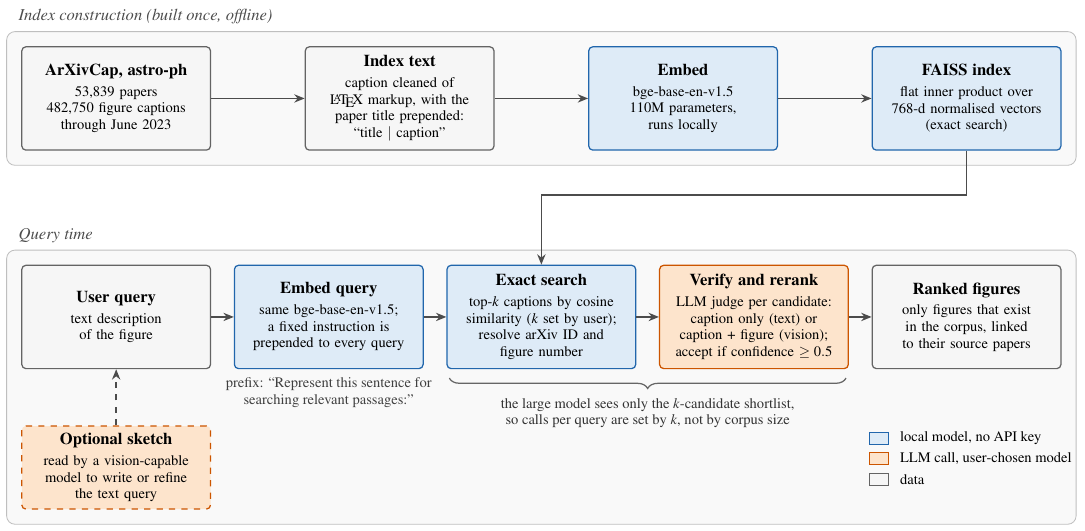}
\caption{The FOTO pipeline. Top: Our index is built once, offline, by cleaning each ArXivCap caption, prepending its paper title, and embedding the result with bge-base-en-v1.5 into an exact FAISS index. Bottom: at query time the user's query is embedded with bge-base-en-v1.5 and the top-$k$ captions are retrieved by cosine similarity. After this, a language model judges each candidate, from its caption alone or from the caption and figure image against the query, and reranks the shortlist. Blue steps run locally with no API key, and orange steps call a language model chosen by the user. Because the language model only sees the $k$ retrieved candidates, its cost per query is set by $k$ rather than by the size of the corpus. A rough sketch can optionally be supplied in place of or alongside the text query (dashed).}
\label{fig:workflow}
\end{figure*}

We began FOTO with an agentic pipeline built on keyword search: a user query was issued to Semantic Scholar \citep{2023arXiv230110140K}, the top-ranked papers were retrieved, their figures were extracted, and each figure was scored by a vision-language model. This approach had two problems. First, recall was poor. Even after expanding the keyword search with domain-informed synonyms and candidate paper titles, the pipeline often failed to retrieve the paper that held the target figure, because a natural-language description of a plot is a poor match for a keyword index. Second, it was expensive: keeping the top 20 candidate papers and extracting their figures left on the order of 200 figures to check, and running a vision-language model on all of them costs both time and money ($\sim$\$0.5 per query).

Replacing the Semantic Scholar search with semantic search over paper abstracts, using \texttt{Pathfinder} \citep{iyer2024pathfinder}, slightly improved the match between the query and the index but did not solve the recall problem well. The pipeline still missed the source paper for a substantial fraction of caption-derived queries, because the signal that matches a figure query lives in the figure caption rather than in the abstract.

This pushed us to index at the level we actually search over which are the individual figures. We index each figure through its caption, embedding the text formed by prepending the paper title to the cleaned caption (Section~\ref{sec:data}) with bge-base-en-v1.5 \citep{xiao2023cpack}, a local model that requires no API key. We chose an openly available model that runs locally so that users need not supply an embedding API key, and a compact one (110M parameters) so that queries can be served on modest CPU-backed hardware at interactive latency. Section~\ref{sec:benchmark} shows this is the strongest of the strategies we tested, and this figure-level caption index is the backbone of FOTO. At query time the user's description is embedded with the same model, using the query instruction prefix the model expects, and matched against the corpus by nearest-neighbor search over a FAISS flat inner-product index \citep{2024arXiv240108281D} of 768-dimensional vectors. Because both corpus and query vectors are normalized, inner products correspond to cosine similarities, and the flat index makes the search exact. The number of top matches carried forward is set by the user. For each match we resolve the arXiv identifier and figure number recorded in the index, retrieve the paper, and extract that figure. Each extracted figure is then passed to a verification model, which judges whether it satisfies the query, using a large language model with an API key supplied by the user. \rev{Users can optionally provide a rough sketch, in which case a vision-capable model is required to interpret it, either to generate a query or to refine a text query given alongside it. We offer sketch input as an option and do not evaluate it quantitatively in this work.} If no sketch is provided and the user selects a text-only model, verification falls back to the retrieved captions rather than the figure images. \rev{ The whole workflow is summarized in Fig. \ref{fig:workflow}. As shown in Section~\ref{vision}, the caption alone is sufficient for verification when the query is well specified, so a vision model is optional for text queries and required only for sketch queries.}

\section{Benchmarks and results}
\label{sec:benchmark}

\subsection{Pilot Study}
\label{sec:pilot}

Before committing to a full-corpus index, we validated the approach on a subsample of the corpus described in Section~\ref{sec:data}. We drew papers at random, stratified to match the publication-year distribution of the full astro-ph slice, until the sample held about 20,000 figures, giving 20{,}122 figures from 2{,}241 papers. We built four figure-level embeddings with \texttt{text-embedding-3-small} (512 dimensions): the caption alone; the caption with the paper title prepended; a rewritten caption, in which DeepSeek~V4~Flash reformulated each caption into a self-contained description using the paper title and abstract for context; and that rewritten caption with the title prepended.

We evaluate retrieval with queries generated from the target figures. \rev{We draw 300 target figures from the subsample and generate one query for each. Nine of these figures have no OCR text in ArXivCap, so we drop them, leaving 291 queries for Table~\ref{tab:pilot_strategy}.} To keep the evaluation conservative, the generator sees only the paper title and optical character recognition text extracted from the figure image available in ArXivCap, never the caption, so no caption wording can leak into the query. This gives a lower-bound estimate of retrieval quality, since a real user often remembers some of the caption's wording. For each query we retrieve the $k$ figures with highest cosine similarity to the query embedding and record the rank $r$ of the target figure. We report recall at $k$, the fraction of queries for which the target figure appears among the top $k$ results. Since each query has exactly one correct target, this is equivalent to an exact hit rate. We also report paper level recall for comparison with other methods. In such cases, if the paper which contains the correct figure is among top $k$ results, we consider that a success.

Table~\ref{tab:pilot_strategy} reports recall at $k$ for the four embeddings. Adding the paper title is what drives retrieval: recall at 20 climbs from 0.491 for the caption alone to 0.914 with the title prepended, and from 0.512 for the rewritten caption to 0.918 once the title is added. Rewriting itself does little. Without the title it lands close to the plain caption (0.512 against 0.491), and with the title it matches plain title-and-caption (0.918 against 0.914). The title carries the figure's identifying signal, so the extra cost of rewriting every caption is not justified. So for the rest of the paper we use title$\,|\,$caption embedding for FOTO's searching.

\begin{table}[t]
\centering
\caption{Pilot retrieval on the 20,122-figure subsample, queries generated from the paper title and figure OCR text, \texttt{text-embedding-3-small}. Recall at $k$ for the exact target figure. Values are over 291 queries, one per target figure, after excluding nine targets with no OCR text.}
\label{tab:pilot_strategy}
\begin{tabular}{lccc}
\hline
index & R@1 & R@5 & R@20 \\
\hline
caption          & 0.155 & 0.296 & 0.491 \\
title\_caption    & 0.309 & 0.639 & 0.914 \\
rewritten        & 0.182 & 0.361 & 0.512 \\
title\_rewritten  & 0.323 & 0.684 & 0.918 \\
\hline
\end{tabular}
\end{table}

Because a user's phrasing and writing style may skew these results, we additionally generate, \rev{for each of the 300 target figures}, five queries in different registers using Claude Haiku~4.5: terse, casual, vague, detailed, and notation (in which symbols are written in a different convention than the caption uses). Table~\ref{tab:pilot_style} reports the title\_caption embedding across these registers. Detailed and terse queries retrieve best, performance is only mildly affected by a change of notation, and, as expected, the largest degradation comes from vague queries.

\begin{table}[t]
\centering
\caption{Pilot retrieval by query register, title\_caption embedding, \texttt{text-embedding-3-small}, 20,122-figure subsample. Recall at $k$ for the exact target figure.}
\label{tab:pilot_style}
\begin{tabular}{lcccccc}
\hline
 & terse & casual & vague & detailed & notation & MEAN \\
\hline
R@1  & 0.640 & 0.427 & 0.140 & 0.813 & 0.527 & 0.509 \\
R@5  & 0.933 & 0.757 & 0.307 & 0.950 & 0.750 & 0.739 \\
R@20 & 0.983 & 0.867 & 0.483 & 0.983 & 0.873 & 0.838 \\
\hline
\end{tabular}
\end{table}

\begin{table*}[t]
\centering
\caption{Embedder comparison on the full corpus, title\_caption embedding, by query register, over the same 500-figure benchmark as Table~\ref{tab:foto-benchmark}. Both models are used as intended: bge-base-en-v1.5 queries carry the model's instruction prefix. Recall at $k$ for the exact target figure.}
\label{tab:embedder}
\begin{tabular}{lccc|ccc}
\hline
 & \multicolumn{3}{c|}{text-embedding-3-small (512d)}
 & \multicolumn{3}{c}{bge-base-en-v1.5 (768d, local)} \\
style & R@1 & R@5 & R@20 & R@1 & R@5 & R@20 \\
\hline
terse    & 0.332 & 0.556 & 0.690 & 0.436 & 0.676 & 0.798 \\
casual   & 0.314 & 0.504 & 0.632 & 0.328 & 0.534 & 0.682 \\
vague    & 0.102 & 0.214 & 0.318 & 0.104 & 0.254 & 0.398 \\
detailed & 0.474 & 0.676 & 0.796 & 0.534 & 0.716 & 0.802 \\
notation & 0.286 & 0.444 & 0.542 & 0.340 & 0.522 & 0.626 \\
MEAN     & 0.302 & 0.479 & 0.596 & 0.348 & 0.540 & 0.661 \\
\hline
\end{tabular}
\end{table*}

Since \texttt{text-embedding-3-small} is served through the OpenAI API, using it to embed queries would tie every search to a paid API call and an OpenAI key. We therefore tested whether a free, openly available model, \texttt{bge-base-en-v1.5}, could build the index instead. Table~\ref{tab:embedder} compares the two on the full corpus over the same 500-figure benchmark used in Table~\ref{tab:foto-benchmark}, broken out by query register, with each model used as intended (the bge queries carry the model's instruction prefix). For this full-corpus benchmark the queries are generated by Claude Haiku~4.5 from the paper title and the figure caption, with an instruction not to copy any phrase longer than four words from the caption (Appendix~\ref{app:prompts}), so they sit closer to the caption wording than the OCR-based pilot queries. The local model is the stronger of the two: its mean recall at 20 is 0.661 against 0.596 for \texttt{text-embedding-3-small}. It leads in four of the five registers, by the most on terse queries in absolute terms (0.798 vs.\ 0.690) and on vague queries in relative terms (0.398 vs.\ 0.318), and the two are tied on detailed queries (0.802 vs.\ 0.796). These full-corpus numbers are lower than the pilot values in Table~\ref{tab:pilot_style}, since they are measured over the entire 482{,}750-caption index rather than the 20,122-figure subsample, and the two experiments also use different query generators, so their values are not directly comparable. The ordering of the registers is nonetheless the same at recall at 1 and 5. Because the free local model outperforms the paid API while removing the per-query cost and key requirement, we build the full-corpus index with \texttt{bge-base-en-v1.5} and drop the OpenAI dependency.

Because the query generators see the paper title, the gain from prepending the title in Table~\ref{tab:pilot_strategy} could in principle come from title words copied into the queries rather than from context a real user would also supply. We test this on the full-corpus benchmark, whose generator sees the title and the caption (Appendix~\ref{app:prompts}). We split titles, captions, and queries into words by lowercasing them and keeping alphanumeric strings of at least two characters, and we remove a trailing ``s'' from words longer than three letters, except those ending in ``ss'', so that singular and plural forms match. We call a word rare if it appears in fewer than 0.1\% of the 53{,}839 paper titles in the index, that is, in at most 53 of them. A query is flagged if it contains a rare word that appears in the title of the target figure's paper but not in the target's caption. Such a word cannot have come from the caption, so it was taken from the title or from the generator's own knowledge. The flagged words are a mix of object and survey names, such as WMAP5, and technical terms that are uncommon in astronomy titles, such as gyrosynchrotron and corotation.

Only 6.3\% of queries (158 of 2,500) are flagged, from 2.6\% in the vague register to 8.2\% in the casual one. Lowering the threshold to 0.01\% of titles flags 0.4 to 3.6\% of queries per register, and raising it to 1\% flags 12 to 21\%. To measure what the flagged words are worth, we delete them from the flagged queries, embed the shortened queries in the same configuration as Table~\ref{tab:foto-benchmark}, and repeat the exact search over the full index, keeping the original queries unchanged everywhere else. Mean recall at 20 falls from 0.661 to 0.646, and by at most 0.026 in any register. Because deletion also removes content a real user would often know, such as the name of the object a figure shows, this is an upper bound on what the copied words are worth. The examples in Appendix~\ref{app:examples} show the same pattern: distinctive title terms rarely appear in the generated queries. The benchmark results therefore do not depend on title words copied into the queries.

\begin{table*}[t]
\centering
\caption{Figure-retrieval recall by query register over 500 sampled figures. FOTO is scored on \emph{figure} hits (target caption in the top~$k$); Pathfinder and Semantic Scholar are scored on the easier \emph{paper} hits (paper containing the target figure in the top~$k$). Semantic Scholar queries use Haiku-extracted keywords; Pathfinder uses the raw query. \rev{Each tool searches its own index: 482,750 captions for FOTO, 385,166 astronomy papers for Pathfinder, and Semantic Scholar's multidisciplinary corpus. With 500 queries per register, the 95\% confidence interval on every entry is at most $\pm0.044$ (Wilson score \citealt{wilson1927}). The smallest gap between FOTO and either baseline, 0.21 for vague queries at R@5, is about 10 standard errors.} \rev{FOTO, in bold, is best in every register at every depth.}}
\label{tab:foto-benchmark}
\begin{tabular}{l | cccc | cccc | cccc}
\toprule
 & \multicolumn{4}{c|}{FOTO (figure)}
 & \multicolumn{4}{c|}{Pathfinder (paper)}
 & \multicolumn{4}{c}{Semantic Scholar (paper)} \\
Style & R@5 & R@10 & R@20 & R@50
      & R@5 & R@10 & R@20 & R@50
      & R@5 & R@10 & R@20 & R@50 \\
\midrule
Terse    & \textbf{0.676} & \textbf{0.742} & \textbf{0.798} & \textbf{0.860} & 0.114 & 0.142 & 0.176 & 0.204 & 0.114 & 0.140 & 0.164 & 0.192 \\
Casual   & \textbf{0.534} & \textbf{0.614} & \textbf{0.682} & \textbf{0.728} & 0.128 & 0.170 & 0.198 & 0.268 & 0.100 & 0.130 & 0.150 & 0.176 \\
Vague    & \textbf{0.254} & \textbf{0.334} & \textbf{0.398} & \textbf{0.470} & 0.042 & 0.048 & 0.058 & 0.094 & 0.044 & 0.066 & 0.092 & 0.106 \\
Detailed & \textbf{0.716} & \textbf{0.766} & \textbf{0.802} & \textbf{0.838} & 0.110 & 0.134 & 0.162 & 0.216 & 0.068 & 0.078 & 0.090 & 0.102 \\
Notation & \textbf{0.522} & \textbf{0.584} & \textbf{0.626} & \textbf{0.684} & 0.070 & 0.090 & 0.116 & 0.148 & 0.050 & 0.066 & 0.076 & 0.082 \\
\midrule
Mean     & \textbf{0.540} & \textbf{0.608} & \textbf{0.661} & \textbf{0.716} & 0.093 & 0.117 & 0.142 & 0.186 & 0.075 & 0.096 & 0.114 & 0.132 \\

\bottomrule
\end{tabular}
\end{table*}

\subsection{FOTO search query benchmark}
\label{comparison}

We use the embedded caption text as the retrieval backbone of FOTO. To compare it with the search tools available to astronomers today, we query two paper-level services on their own indexes: (i) Pathfinder \citep{iyer2024pathfinder}, a semantic search tool over 385,166 peer-reviewed astronomy papers from ADS, and (ii) Semantic Scholar \citep{2023arXiv230110140K}, which indexes more than 200 million papers across the natural, physical, and social sciences and whose relevance search matches keywords against titles, abstracts, and author names, so we query it with keywords extracted from each query by Claude Haiku~4.5. For Pathfinder we reproduce its dense retrieval by embedding each query with \texttt{text-embedding-3-small}, the same model Pathfinder uses for its corpus, and the 385,166 papers in that corpus are ranked by cosine similarity to the query. For both baselines we score a hit if the paper containing the target figure appears among the top-$k$ retrieved papers, whereas for FOTO we require the target figure itself to appear among the top-$k$ retrieved captions. This criterion favors the baselines: because each paper contains $8$-$9$ figures on average, locating the correct paper is a substantially easier task than locating the correct figure. We test across five query registers over the same 500 sampled figures.

\rev{These queries are generated from the paper title and the figure caption (Section~\ref{sec:pilot}), which favors a caption index such as FOTO's over the abstract-level indexes of the baselines. Although the generator was told not to copy phrases longer than four words, some queries share a run of five or more consecutive words with their target caption: 1.6\% of terse, 5.8\% of casual, 1.4\% of vague, 30.6\% of detailed, and 7.6\% of notation queries, and 9.4\% overall. Some of these shared spans are object names that a user searching for the figure would also use, such as HD~19994 or EGO~G27.97$-$0.47. Restricting the comparison to the 2{,}265 queries that share no phrases longer than four words lowers FOTO's mean recall at 20 from 0.661 to 0.645, while Pathfinder and Semantic Scholar are essentially unchanged at 0.143 and 0.119, so the copying does not substantially alter FOTO's performance.}

FOTO outperforms both baselines by a wide margin in every register, despite being held to the stricter figure-level criterion. At R@20, FOTO recovers the target figure for $40$ -$80\%$ of queries depending on register, compared with $6$-$20\%$ for Pathfinder and $8$-$16\%$ for Semantic Scholar at the (easier) paper level. The margin is largest for well-specified registers (\emph{detailed}, \emph{terse}) and narrows but persists for under-specified ones; the \emph{vague} register remains the hardest for all three methods, indicating headroom rather than a ceiling. \rev{Against Pathfinder, the most direct comparison, FOTO finds the exact figure about 4-5 times as often as Pathfinder finds the right paper (mean recall at 20 of 0.661 against 0.142). Because each tool searches its own index, these numbers describe what a researcher gets from each tool today rather than a comparison of retrieval methods on a shared corpus.}


\begin{table*}[t]
\centering
\caption{\rev{Verification at retrieval depth 50, acceptance confidence $\geq 0.5$. Every row reranks the same 50 candidates per query. R@1 and R@5 are the fraction of queries whose target figure lands in the top 1 and top 5 after reranking, and real-accept is how often the judge accepts the true figure. Vision judge see the figure image with its caption, text judges see the caption alone. Recall is computed over the queries whose target reached the pool. The last row is dense retrieval alone on the same candidates. Superscripts and subscripts give the 95\% interval for each value, from the number of hits. Best judge in each column in bold.}}
\label{tab:vision}
\begin{tabular}{l|ccc|ccc}
\hline
 & \multicolumn{3}{c|}{detailed} & \multicolumn{3}{c}{vague} \\
judge & R@1 & R@5 & real-accept & R@1 & R@5 & real-accept \\
\hline
vision (Sonnet) & 0.708$^{+0.143}_{-0.200}$ & 0.750$^{+0.130}_{-0.199}$ & 0.750$^{+0.130}_{-0.199}$ & 0.429$^{+0.246}_{-0.215}$ & \textbf{0.929$^{+0.059}_{-0.243}$} & \textbf{1.000$^{\phantom{+0.000}}_{-0.215}$} \\
text (Sonnet) & \textbf{0.875$^{+0.082}_{-0.185}$} & 0.917$^{+0.060}_{-0.175}$ & 0.958$^{+0.034}_{-0.161}$ & 0.357$^{+0.255}_{-0.194}$ & 0.857$^{+0.103}_{-0.257}$ & \textbf{1.000$^{\phantom{+0.000}}_{-0.215}$} \\
text (DeepSeek) & 0.750$^{+0.130}_{-0.199}$ & 0.792$^{+0.116}_{-0.196}$ & 0.875$^{+0.082}_{-0.185}$ & 0.214$^{+0.262}_{-0.139}$ & 0.643$^{+0.194}_{-0.255}$ & 0.857$^{+0.103}_{-0.257}$ \\

text (Kimi K3) & 0.833$^{+0.100}_{-0.192}$ & \textbf{1.000$^{\phantom{+0.000}}_{-0.138}$} & \textbf{1.000$^{\phantom{+0.000}}_{-0.138}$} & \textbf{0.571$^{+0.215}_{-0.246}$} & 0.857$^{+0.103}_{-0.257}$ & \textbf{1.000$^{\phantom{+0.000}}_{-0.215}$} \\

text (Kimi K2.6) & \textbf{0.875$^{+0.082}_{-0.185}$} & \textbf{1.000$^{\phantom{+0.000}}_{-0.138}$} & \textbf{1.000$^{\phantom{+0.000}}_{-0.138}$} & 0.500$^{+0.232}_{-0.232}$ & 0.857$^{+0.103}_{-0.257}$ & \textbf{1.000$^{\phantom{+0.000}}_{-0.215}$} \\
\hline
retrieval only & 0.667$^{+0.154}_{-0.200}$ & 0.917$^{+0.060}_{-0.175}$ &  & 0.071$^{+0.243}_{-0.059}$ & 0.571$^{+0.215}_{-0.246}$ &  \\
\hline
\end{tabular}
\end{table*}

\subsection{Verification benchmark}
\label{vision}

FOTO adds a verification stage that re-examines the retrieved candidates and decides which satisfy the query. We benchmark it on 60 queries across the two extreme registers, 30 detailed and 30 vague, reusing the queries from Section~\ref{comparison}. For each query we retrieve the top 50 candidates and rerank them with an LLM judge under two input conditions: vision, where the judge sees the figure image with its caption, and text, where it sees the caption alone. The queries in this case are where the target reaches the top 50 since no judge can recover a figure that was never retrieved. We compare a frontier model (\texttt{claude-sonnet-4-6}) with openly available models (Kimi K3, Kimi K2.6, and DeepSeek). All judges share a single prompt (Appendix~\ref{app:prompts}) which holds a candidate to the specificity of the query. A candidate is accepted at match confidence $\geq 0.5$.

Verification improves the ranking in both registers (Table~\ref{tab:vision}). Dense retrieval places the target first for 0.667 of detailed queries and 0.071 of vague ones, and the best judges raise these to 0.875 and 0.571. The gain is largest where retrieval is weakest. On vague queries the target is usually somewhere in the pool but rarely at the top, and reranking lifts recall at 5 from 0.571 to 0.929. On detailed queries retrieval already places the target in the top 5 for 0.917 of queries, so there is little headroom, although both Kimi text judges still reach 1.000.

The caption alone suffices when the query is well specified. On detailed queries the text condition is at least as good as the vision. On vague queries the ordering flips for Sonnet, where vision leads at recall 5. 

Because the benchmark queries were generated with Claude Haiku~4.5, a judge from the same model family could in principle be favored by shared phrasing rather than by genuine relevance. To test this we include judges from two other developers. Kimi K2.6 matches or exceeds the Sonnet text condition on every metric in Table~\ref{tab:vision}, and Kimi K3 gives the highest vague recall at 1 (0.571 against 0.429 for the best Sonnet condition), so the verification gains do not depend on the judge sharing a model family with the query generator. DeepSeek is the weakest text judge in both registers. These comparisons rest on small samples. With 24 detailed and 14 vague queries a single query moves recall by 0.042 and 0.071, the 95\% Wilson interval for a recall at 1 of 0.875 on detailed queries spans 0.69 to 0.96, and the differences between the top judges amount to one or two queries. We conclude that judges performances are not unfairly skewed by which model was used to generate queries.

\subsection{Cost and latency}
\label{sec:cost}

Retrieval runs locally, costs nothing per query, and is fast: embedding the query and searching the flat index over 482{,}750 captions takes well under a second on a CPU. The time and cost of a query are instead set by the steps that follow, fetching each candidate figure and verifying it with a language model, and both grow with the number of candidates $k$ rather than with the size of the corpus. With caption-only verification, which Section~\ref{vision} shows is sufficient for well-specified queries, judging one candidate with \texttt{claude-sonnet-4-6} costs about \$0.003 to \$0.005, so a query verified at $k=20$ costs about \$0.06 to \$0.10. DeepSeek is more than ten times cheaper per candidate. Retrieval alone returns its ranked candidates in well under a second, while a complete query with verification at $k=20$ currently takes about 3-4 minutes, most of it spent fetching the candidate figures and waiting on the judge calls.

\section{FOTO: Data and code availability}
\label{sec:availability}
The webapp for FOTO is hosted at \href{https://huggingface.co/spaces/htohfa/foto}{https://huggingface.co/spaces/htohfa/foto}. The dataset is available at: \href{https://huggingface.co/datasets/htohfa/foto-index}{https://huggingface.co/datasets \\ /htohfa/foto-index}. The open source code is also available on github (\href{https://github.com/htohfa/Figure-Finder/}{https://github.com/htohfa/Figure-Finder/}) for recreating all the results and running the webapp.  


\section{Conclusion}
\label{sec:conclusion}

We presented FOTO, a caption-first tool for retrieving published astronomy figures from a text description. Our early attempts at paper-level retrieval, first via keyword search using Semantic Scholar and then via semantic search over abstracts with Pathfinder, both fell short because the signal that matches a figure query lives in the caption, not in the abstract or the paper's keywords. Indexing at the figure level directly fixes this. Embedding paper title  with the raw caption raised recall at 20 top queries from 0.491 for the caption alone to 0.914 in our pilot study with 20,122 figures (Table 1). At full corpus scale, our locally hosted \texttt{bge-base-en-v1.5} model reached a mean recall at 20 of 0.661 against 0.596 for the paid \texttt{text-embedding-3-small}, with the largest relative gain on vague queries. \rev{We also compare FOTO with the paper-level search tools available to astronomers today.} FOTO recovers the target figure for 40-80\% of queries at R@20 depending on register, compared with 6-20\% for Pathfinder and 8-16\% for Semantic Scholar at the paper-level task (Table 4).

\rev{After finding the top matches with our embedding, FOTO verifies them with a user chosen language model. Verification is the component that turns the shortlist into a ranked answer: reranking the retrieved pool raises recall at 1 from 0.667 to 0.875 on detailed queries and from 0.071 to 0.571 on vague ones (Table~\ref{tab:vision}). The caption alone suffices when the query is well specified, and the image mainly matters for sketch queries where no text description exists. Openly available judges perform at least as well as a paid model and the gains do not come from the judge sharing a model family with the query generator.} Vague, underspecified queries remain the clear limitation across every stage of the pipeline, and closing that gap through fine-tuning the retrieval is the natural next step. All results in this paper use the astro-ph papers in ArXivCap, which end in June 2023 and include only papers recorded as published. \rev{We are extending the index beyond June 2023 to include recent papers and preprints, and we plan to update the hosted index and web application regularly as new papers appear.}


\bibliography{references}

@article{kurtz2000ads,
  author  = {Kurtz, Michael J. and Eichhorn, Guenther and Accomazzi, Alberto and Grant, Carolyn S. and Murray, Stephen S. and Watson, Joyce M.},
  title   = {The {NASA} {Astrophysics} {Data} {System}: Overview},
  journal = {Astronomy and Astrophysics Supplement Series},
  volume  = {143},
  pages   = {41--59},
  year    = {2000},
  doi     = {10.1051/aas:2000170}
}

@article{iyer2024pathfinder,
   title={pathfinder: A Semantic Framework for Literature Review and Knowledge Discovery in Astronomy},
   volume={275},
   ISSN={1538-4365},
   url={http://dx.doi.org/10.3847/1538-4365/ad7c43},
   DOI={10.3847/1538-4365/ad7c43},
   number={2},
   journal={The Astrophysical Journal Supplement Series},
   publisher={American Astronomical Society},
   author={Iyer, Kartheik G. and Yunus, Mikaeel and O’Neill, Charles and Ye, Christine and Hyk, Alina and McCormick, Kiera and Ciucă, Ioana and Wu, John F. and Accomazzi, Alberto and Astarita, Simone and Chakrabarty, Rishabh and Cranney, Jesse and Field, Anjalie and Ghosal, Tirthankar and Ginolfi, Michele and Huertas-Company, Marc and Jabłońska, Maja and Kruk, Sandor and Liu, Huiling and Marchidan, Gabriel and Mistry, Rohit and Naiman, J. P. and Peek, J. E. G. and Polimera, Mugdha and Rodríguez Méndez, Sergio J. and Schawinski, Kevin and Sharma, Sanjib and Smith, Michael J. and Ting, Yuan-Sen and Walmsley, Mike and },
   year={2024},
   month=Nov, pages={38} }

@inproceedings{li2024multimodalarxiv,
  author    = {Li, Lei and Wang, Yuqi and Xu, Runxin and Wang, Peiyi and Feng, Xiachong and Kong, Lingpeng and Liu, Qi},
  title     = {Multimodal {ArXiv}: A Dataset for Improving Scientific Comprehension of Large Vision-Language Models},
  booktitle = {Proceedings of the 62nd Annual Meeting of the Association for Computational Linguistics},
  year      = {2024},
  eprint    = {2403.00231},
  archivePrefix = {arXiv}
}

@article{wu2024scimmir,
  author  = {Wu, Siwei and Li, Yizhi and Zhu, Kang and Zhang, Ge and Liang, Yiming and Ma, Kaijing and Xiao, Chenghao and Zhang, Haoran and Yang, Bohao and Chen, Wenhu and others},
  title   = {{SciMMIR}: Benchmarking Scientific Multi-modal Information Retrieval},
  journal = {arXiv e-prints},
  eprint  = {2401.13478},
  archivePrefix = {arXiv},
  year    = {2024}
}

@inproceedings{roberts2024scifibench,
  author    = {Roberts, Jonathan and Han, Kai and Houlsby, Neil and Albanie, Samuel},
  title     = {{SciFIBench}: Benchmarking Large Multimodal Models for Scientific Figure Interpretation},
  booktitle = {Advances in Neural Information Processing Systems},
  volume    = {37},
  year      = {2024},
  eprint    = {2405.08807},
  archivePrefix = {arXiv}
}

@article{xiao2023cpack,
  author  = {Xiao, Shitao and Liu, Zheng and Zhang, Peitian and Muennighoff, Niklas},
  title   = {{C-Pack}: Packaged Resources To Advance General Chinese Embedding},
  journal = {arXiv e-prints},
  eprint  = {2309.07597},
  archivePrefix = {arXiv},
  year    = {2023},
  note    = {Canonical citation for the BGE embedding and reranker models}
}

@inproceedings{clark2016pdffigures,
  author    = {Clark, Christopher and Divvala, Santosh},
  title     = {{PDFFigures} 2.0: Mining Figures from Research Papers},
  booktitle = {Proceedings of the 16th ACM/IEEE-CS Joint Conference on Digital Libraries (JCDL)},
  year      = {2016},
  doi       = {10.1145/2910896.2910904}
}

@article{huang2023hallucination,
  author  = {Huang, Lei and Yu, Weijiang and Ma, Weitao and Zhong, Weihong and Feng, Zhangyin and Wang, Haotian and Chen, Qianglong and Peng, Weihua and Feng, Xiaocheng and Qin, Bing and Liu, Ting},
  title   = {A Survey on Hallucination in Large Language Models: Principles, Taxonomy, Challenges, and Open Questions},
  journal = {arXiv e-prints},
  eprint  = {2311.05232},
  archivePrefix = {arXiv},
  year    = {2023},
  note    = {Published in ACM Transactions on Information Systems (2025)}
}

@INPROCEEDINGS{grezes2021astrobert,
       author = {{Grezes}, F. and {Blanco-Cuaresma}, S. and {Accomazzi}, A. and {Kurtz}, M.~J. and {Shapurian}, G. and {Henneken}, E. and {Grant}, C.~S. and {Thompson}, D.~M. and {Chyla}, R. and {McDonald}, S. and {Hostetler}, T.~W. and {Templeton}, M.~R. and {Lockhart}, K.~E. and {Martinovic}, N. and {Chen}, S. and {Tanner}, C. and {Protopapas}, P.},
        title = "{Building astroBERT, a Language Model for Astronomy \& Astrophysics}",
    booktitle = {Astronomical Data Analysis Software and Systems XXXI},
         year = 2024,
       editor = {{Hugo}, B.~V. and {Van Rooyen}, R. and {Smirnov}, O.~M.},
       series = {Astronomical Society of the Pacific Conference Series},
       volume = {535},
        month = may,
        pages = {119},
          doi = {10.48550/arXiv.2112.00590},
archivePrefix = {arXiv},
       eprint = {2112.00590},
 primaryClass = {cs.CL},
       adsurl = {https://ui.adsabs.harvard.edu/abs/2024ASPC..535..119G}
}

@ARTICLE{nguyen2023astrollama,
       author = {{Dung Nguyen}, Tuan and {Ting}, Yuan-Sen and {Ciuc{\u{a}}}, Ioana and {O'Neill}, Charlie and {Sun}, Ze-Chang and {Jab{\l}o{\'n}ska}, Maja and {Kruk}, Sandor and {Perkowski}, Ernest and {Miller}, Jack and {Li}, Jason and {Peek}, Josh and {Iyer}, Kartheik and {R{\'o}{\.z}a{\'n}ski}, Tomasz and {Khetarpal}, Pranav and {Zaman}, Sharaf and {Brodrick}, David and {Rodr{\'\i}guez M{\'e}ndez}, Sergio J. and {Bui}, Thang and {Goodman}, Alyssa and {Accomazzi}, Alberto and {Naiman}, Jill and {Cranney}, Jesse and {Schawinski}, Kevin and {UniverseTBD}},
        title = "{AstroLLaMA: Towards Specialized Foundation Models in Astronomy}",
      journal = {arXiv e-prints},
         year = 2023,
        month = sep,
          eid = {arXiv:2309.06126},
        pages = {arXiv:2309.06126},
          doi = {10.48550/arXiv.2309.06126},
archivePrefix = {arXiv},
       eprint = {2309.06126},
 primaryClass = {astro-ph.IM},
       adsurl = {https://ui.adsabs.harvard.edu/abs/2023arXiv230906126D}
}

@ARTICLE{2019arXiv190810084R,
       author = {{Reimers}, Nils and {Gurevych}, Iryna},
        title = "{Sentence-BERT: Sentence Embeddings using Siamese BERT-Networks}",
      journal = {arXiv e-prints},
         year = 2019,
        month = aug,
          eid = {arXiv:1908.10084},
        pages = {arXiv:1908.10084},
          doi = {10.48550/arXiv.1908.10084},
archivePrefix = {arXiv},
       eprint = {1908.10084},
 primaryClass = {cs.CL},
       adsurl = {https://ui.adsabs.harvard.edu/abs/2019arXiv190810084R}
}

@ARTICLE{2026arXiv260212303L,
       author = {{Lewis}, Rommulus Francis and {Shah}, Hetansh and {Alfred}, Amruth},
        title = "{Astrophysics Wrapped 2025: Year-in-Review of Every Astrophysics arXiv Paper from 2025}",
      journal = {arXiv e-prints},
         year = 2026,
        month = feb,
          eid = {arXiv:2602.12303},
        pages = {arXiv:2602.12303},
          doi = {10.48550/arXiv.2602.12303},
archivePrefix = {arXiv},
       eprint = {2602.12303},
 primaryClass = {astro-ph.IM},
       adsurl = {https://ui.adsabs.harvard.edu/abs/2026arXiv260212303L}
}

@ARTICLE{2023arXiv230110140K,
       author = {{Kinney}, Rodney and {Anastasiades}, Chloe and {Authur}, Russell and {Beltagy}, Iz and {Bragg}, Jonathan and {Buraczynski}, Alexandra and {Cachola}, Isabel and {Candra}, Stefan and {Chandrasekhar}, Yoganand and {Cohan}, Arman and {Crawford}, Miles and {Downey}, Doug and {Dunkelberger}, Jason and {Etzioni}, Oren and {Evans}, Rob and {Feldman}, Sergey and {Gorney}, Joseph and {Graham}, David and {Hu}, Fangzhou and {Huff}, Regan and {King}, Daniel and {Kohlmeier}, Sebastian and {Kuehl}, Bailey and {Langan}, Michael and {Lin}, Daniel and {Liu}, Haokun and {Lo}, Kyle and {Lochner}, Jaron and {MacMillan}, Kelsey and {Murray}, Tyler and {Newell}, Chris and {Rao}, Smita and {Rohatgi}, Shaurya and {Sayre}, Paul and {Shen}, Zejiang and {Singh}, Amanpreet and {Soldaini}, Luca and {Subramanian}, Shivashankar and {Tanaka}, Amber and {Wade}, Alex D. and {Wagner}, Linda and {Wang}, Lucy Lu and {Wilhelm}, Chris and {Wu}, Caroline and {Yang}, Jiangjiang and {Zamarron}, Angele and {Van Zuylen}, Madeleine and {Weld}, Daniel S.},
        title = "{The Semantic Scholar Open Data Platform}",
      journal = {arXiv e-prints},
         year = 2023,
        month = jan,
          eid = {arXiv:2301.10140},
        pages = {arXiv:2301.10140},
          doi = {10.48550/arXiv.2301.10140},
archivePrefix = {arXiv},
       eprint = {2301.10140},
 primaryClass = {cs.DL},
       adsurl = {https://ui.adsabs.harvard.edu/abs/2023arXiv230110140K}
}

@ARTICLE{2021arXiv210300020R,
       author = {{Radford}, Alec and {Kim}, Jong Wook and {Hallacy}, Chris and {Ramesh}, Aditya and {Goh}, Gabriel and {Agarwal}, Sandhini and {Sastry}, Girish and {Askell}, Amanda and {Mishkin}, Pamela and {Clark}, Jack and {Krueger}, Gretchen and {Sutskever}, Ilya},
        title = "{Learning Transferable Visual Models From Natural Language Supervision}",
      journal = {arXiv e-prints},
         year = 2021,
        month = feb,
          eid = {arXiv:2103.00020},
        pages = {arXiv:2103.00020},
          doi = {10.48550/arXiv.2103.00020},
archivePrefix = {arXiv},
       eprint = {2103.00020},
 primaryClass = {cs.CV},
       adsurl = {https://ui.adsabs.harvard.edu/abs/2021arXiv210300020R}
}

@ARTICLE{2024arXiv240108281D,
       author = {{Douze}, Matthijs and {Guzhva}, Alexandr and {Deng}, Chengqi and {Johnson}, Jeff and {Szilvasy}, Gergely and {Mazar{\'e}}, Pierre-Emmanuel and {Lomeli}, Maria and {Hosseini}, Lucas and {J{\'e}gou}, Herv{\'e}},
        title = "{The Faiss library}",
      journal = {arXiv e-prints},
         year = 2024,
        month = jan,
          eid = {arXiv:2401.08281},
        pages = {arXiv:2401.08281},
          doi = {10.48550/arXiv.2401.08281},
archivePrefix = {arXiv},
       eprint = {2401.08281},
 primaryClass = {stat.ML},
       adsurl = {https://ui.adsabs.harvard.edu/abs/2024arXiv240108281D}
}

@misc{aie2013,
  author       = {{American Astronomical Society}},
  title        = {{Astronomy Image Explorer}},
  year         = {2013},
  howpublished = {\url{https://www.astroexplorer.org/}},
  note         = {Accessed October 2026}
}

@article{demnerfushman2012openi,
  author  = {Demner-Fushman, Dina and Antani, Sameer K. and Simpson, Matthew and Thoma, George R.},
  title   = {Design and Development of a Multimodal Biomedical Information Retrieval System},
  journal = {Journal of Computing Science and Engineering},
  year    = {2012},
  volume  = {6},
  number  = {2},
  pages   = {168--177},
  doi     = {10.5626/JCSE.2012.6.2.168}
}

@article{wilson1927,
  author  = {Wilson, Edwin B.},
  title   = {Probable Inference, the Law of Succession, and Statistical Inference},
  journal = {Journal of the American Statistical Association},
  year    = {1927},
  volume  = {22},
  number  = {158},
  pages   = {209--212}
}

\appendix
\section{LLM Prompts used}
\label{app:prompts}

Here we list the prompts used in this work. Each prompt is sent as a single user message with no system prompt and is reproduced verbatim, except that long lines are wrapped for display. Placeholders in braces, such as \texttt{\{title\}}, are replaced with the corresponding value before the prompt is sent.

\subsection{Benchmark queries}
\label{app:prompts-queries}

For each randomly sampled target figure, we ask the model to write one query in each of the five registers used throughout the paper (terse, casual, vague, detailed, and notation). The model is given the paper title and the figure caption.

\medskip
\textbf{Model and settings:} \texttt{claude-haiku-4-5-20251001}, at most 2000 output tokens, no temperature set (API default).
\medskip

You are generating benchmark search queries for a figure retrieval system. Below is the title of an astronomy paper and the caption of one figure from it. Write five queries a researcher might type when looking for this exact figure, WITHOUT having the caption in front of them: \\
1. ``terse": short expert phrasing, 4-10 words, standard jargon. \\
2. ``casual": conversational, as if asking a colleague, one sentence. \\
3. ``vague": the researcher half-remembers the figure. Keep it faithful to the figure's actual content but drop specifics (no survey names, no exact quantities); 5-15 words. \\
4. ``detailed": two sentences describing axes, curves, and what is compared. \\
5. ``notation": uses symbols/notation an expert would use (e.g. $sigma_8$, $z\sim 2$, $P(k)$).

Do not copy phrases longer than 4 words from the caption. Respond with ONLY a JSON object: {``terse": ``...", ``casual": ``...", ``vague": ``...", ``detailed": ``...", ``notation": ``..."}

Title: \{title\}

Caption: \{caption\}

\subsection{Verification Judge}

\textbf{Model and settings:} Sonnet claude-sonnet-4-6, max tokens 250; DeepSeek judge:deepseek-chat, max tokens 250; Kimi kimi-k3 with reasoning effort low and kimi-k2.6 with thinking disabled, JSON mode, max tokens 1000 (raised up because replies ran out). Accept at match = true and confidence $>=$ 0.5.
\medskip

A researcher is searching for a published scientific figure with this query:

``{query}"

Caption: {caption}

Judge the candidate against what the query actually specifies. If the query is broad or vague, any figure genuinely showing what it describes counts as a match; do not demand details the query never mentions. If the query specifies particulars (quantities, axes, plot type, what is compared), the figure must show them. Science match is the dominant factor: a figure on the wrong scientific topic should never score above 0.4.

\noindent Respond JSON only:\\
\{``match'': true$\,|\,$false, ``confidence'': 0.0-1.0, ``what\_is\_plotted'': ``$<$one sentence$>$'', ``reason'': ``$<$one sentence$>$''\}\\[4pt]
vision: Below is the candidate figure image together with its caption.\\
text: Only the figure's caption is available (no image).\\
sketch (appended, if given): The researcher also provided a rough sketch of the figure they remember (first image); weigh structural similarity to it.

\section{Example Queries}\label{app:examples}

Here we show example queries for twenty randomly sampled figures from the full-corpus benchmark. For each figure, Claude Haiku~4.5 was given the paper title and the figure caption, as described in Section~\ref{sec:pilot} and Appendix~\ref{app:prompts}, and generated one query in each of the five registers (terse, casual, vague, detailed, and notation). Titles and captions are shown as they appear in ArXivCap, including the \texttt{<cit.>} and \texttt{<ref>} tokens that replace citations and cross references. The last column gives the rank at which the target figure is retrieved, so a rank of 1 means it is the top result, and N/A means it was not retrieved within the top 100 results.

\begin{fotoexample}{2304.02727}{1}
\extitle{Fine-structure transitions of Si and S induced by collisions with atomic hydrogen}
\excaption{Potential energies for the SiH system as functions of the internuclear distance $R$, labeled from the top down: b$^4\Pi$ (blue dashed line), B$^2\Sigma^-$ (green dotted line) , a$^4\Sigma^-$ (orange solid line), X$^2\Pi$ (red dashdot line).}
\begin{querytable}
\qrow{terse}{SiH potential energy curves internuclear distance}{1}
\qrow{casual}{Can you show me the potential energy diagram for the SiH system as a function of distance?}{2}
\qrow{vague}{potential energy curves for silicon hydride system plotted against separation distance}{1}
\qrow{detailed}{Plot showing potential energies for the SiH system versus internuclear distance R, with multiple electronic states displayed as different colored curves including b4$\Pi$, B2$\Sigma$-, and a4$\Sigma$- states.}{1}
\qrow{notation}{$V(R)$ for SiH: b$^4\Pi$, B$^2\Sigma^-$, a$^4\Sigma^-$ states}{2}
\end{querytable}
\end{fotoexample}

\begin{fotoexample}{2202.05141}{24}
\extitle{An observational synthesis of the Taurid meteor complex}
\excaption{Radiant drift of Taurid streams as a function of solar longitude of the right ascension, declination, sun-centered ecliptic latitude and longitude (bottom panels). Shown are the STA as measured by CMOR (wavelet analysis, in blue), CAMS and the IAU MDC \texttt{<cit.>}. The STA identified within the CAMS database between 2010 and 2016 are represented in grey, while the shower's sub-components identified by \texttt{<cit.>} are identified in green. The solid curves illustrate the result of a linear or quadratic fit of each data set, line color corresponding to symbol colors. Fit results can be found in Table 2.}
\begin{querytable}
\qrow{terse}{Taurid radiant drift solar longitude parameters}{1}
\qrow{casual}{Can you show me how the Taurid radiant position changes with solar longitude across different coordinate systems?}{1}
\qrow{vague}{Figure showing how meteor stream radiant drifts with solar longitude in multiple coordinate systems}{2}
\qrow{detailed}{Four-panel figure displaying radiant drift of Taurid streams as functions of solar longitude, with right ascension, declination, ecliptic latitude, and ecliptic longitude shown separately on bottom panels with CMO measurements.}{1}
\qrow{notation}{$\alpha$, $\delta$, $\beta_{\rm ecl}$, $\lambda_{\rm ecl}$ vs $L_\odot$ for Taurid radiant displacement}{32}
\end{querytable}
\end{fotoexample}

\begin{fotoexample}{0907.5424}{3}
\extitle{TASI Lectures on Inflation}
\excaption{Conformal diagram of inflationary cosmology. Inflation extends conformal time to negative values! The end of inflation creates an ``apparent'' Big Bang at $\tau = 0$. There is, however, no singularity at $\tau = 0$ and the light cones intersect at an earlier time if inflation lasts for at least 60 $e$-folds.}
\begin{querytable}
\qrow{terse}{conformal diagram inflation cosmology timeline}{2}
\qrow{casual}{Do you have that conformal diagram showing how inflation extends time backwards to a kind of artificial Big Bang?}{1}
\qrow{vague}{diagram showing time evolution during and after inflation with special point at end}{12}
\qrow{detailed}{Conformal diagram plotting conformal time on horizontal axis, showing how inflation extends into negative time values and creates an apparent Big Bang singularity at the transition point.}{1}
\qrow{notation}{conformal time $\tau$ diagram inflationary epoch, $\tau = 0$ transition}{2}
\end{querytable}
\end{fotoexample}

\begin{fotoexample}{2110.00034}{9}
\extitle{Bent it like FRs: extended radio AGN in the COSMOS field and their large-scale environment}
\excaption{Bent angle in degrees of COSMOS FR sources at 3 GHz versus redshift. For reference, the large-scale environments we use in the analysis cover the redshift range up to $z \sim 3$, depending on the probe.}
\begin{querytable}
\qrow{terse}{FR bent angle redshift COSMOS 3 GHz}{2}
\qrow{casual}{Can you show me the plot of how bent these radio sources are versus their redshift?}{6}
\qrow{vague}{bent angle versus redshift for radio sources at different distances}{3}
\qrow{detailed}{Scatter plot showing bent angle in degrees on the y-axis versus redshift on the x-axis for COSMOS FR sources observed at 3 GHz, with redshift range extending to $z \sim 3$.}{2}
\qrow{notation}{bent angle (degrees) vs $z$, 3 GHz COSMOS FRs, $z \leq 3$}{2}
\end{querytable}
\end{fotoexample}

\begin{fotoexample}{1410.1181}{5}
\extitle{Collision-free motion planning for fiber positioner robots: discretization of velocity profiles}
\excaption{(A) A position profile with the safety margin shown in grey. In the first step, we draw a line from $t_0$ to $t_f$ on the profile. This line is entirely in the safety zone. By changing the position profile from the continuous curve to the single line, still collision avoidance is guaranteed. (B,C,D) A single line can not provide a collision free trajectory for the positioner. (C)Therefore, we find a middle point between $t_0$ and $t_f$ as a breaking point. But the line is not still entirely in the safety margin, so we need to repeat the last step, till we get line segments entirely in the safety margin (D).}
\begin{querytable}
\qrow{terse}{position profile safety margin line}{1}
\qrow{casual}{Can you show me that figure with the position profile and the grey safety zone around it?}{1}
\qrow{vague}{graph showing a trajectory with a safety buffer region highlighted}{2}
\qrow{detailed}{Figure showing a position profile over time with a grey safety margin region, including a line drawn from initial to final time that remains within the safe zone.}{1}
\qrow{notation}{$q(t)$ profile with safety margin visualization and feasible path in configuration space}{1}
\end{querytable}
\end{fotoexample}

\begin{fotoexample}{0709.3113}{1}
\extitle{Point Sources from a Spitzer IRAC Survey of the Galactic Center}
\excaption{Distribution of point sources with Galactic longitude for the three defined brightness ranges, as indicated. Only the sources located within $-1.0 \leq l \leq 1.0$ and $-0.7 \leq b \leq 0.7$ are included in the determination of the number density distribution. Symbols are the same as in Figure \texttt{<ref>}.}
\begin{querytable}
\qrow{terse}{Point source distribution Galactic longitude brightness ranges}{1}
\qrow{casual}{Can you show me the histogram of how point sources are distributed across Galactic longitude in different brightness categories?}{2}
\qrow{vague}{Distribution of sources by longitude split into brightness categories near Galactic center}{1}
\qrow{detailed}{A histogram showing the distribution of point sources as a function of Galactic longitude, with separate curves or bars for three different brightness ranges. The x-axis shows Galactic longitude and the y-axis shows the number of sources.}{10}
\qrow{notation}{$N(l)$ for three flux bins, $l \in [-1.0, 1.0]^\circ$, $b \in [-0.7, 0.7]^\circ$}{N/A}
\end{querytable}
\end{fotoexample}

\begin{fotoexample}{0807.2147}{1}
\extitle{UHECR bounds on Lorentz violation in the photon sector}
\excaption{Left: classical spacetime-foam manifold. Right: Feynman diagram for vacuum Cherenkov radiation.}
\begin{querytable}
\qrow{terse}{spacetime foam manifold Feynman diagram vacuum Cherenkov}{1}
\qrow{casual}{Do you have that figure showing the spacetime foam manifold next to the Feynman diagram for vacuum Cherenkov radiation?}{1}
\qrow{vague}{comparison of theoretical spacetime structure and particle interaction diagram for radiation process}{N/A}
\qrow{detailed}{Left panel displays a classical spacetime-foam manifold structure. Right panel shows a corresponding Feynman diagram illustrating vacuum Cherenkov radiation.}{1}
\qrow{notation}{spacetime foam manifold vs.\ Feynman diagram for vacuum Cherenkov emission}{1}
\end{querytable}
\end{fotoexample}

\begin{fotoexample}{1906.00994}{16}
\extitle{Probing the existence of very massive first stars}
\excaption{Iron abundances of an ISM imprinted by both PISN of mass ${\rm m_{\tt PISN}}$ and the subsequent generation of Pop II stars. The results refer to the model $=10^{-3}$ (Fig.~\texttt{<ref>} and text), and assume ${\rm f_{PISN}=0.9}$ ({left}) and ${\rm f_{PISN}=0.5}$ ({right}). Different symbols show the expected values at different times since the onset of Pop II star-formation (see labels).}
\begin{querytable}
\qrow{terse}{Iron abundances PISN Pop II stars comparison}{1}
\qrow{casual}{Can you show me that plot with iron abundances from the first stars and second generation stars?}{34}
\qrow{vague}{Iron abundance predictions from massive first stars and their descendants}{2}
\qrow{detailed}{Figure showing iron abundances in the interstellar medium comparing contributions from pair-instability supernovae of very massive stars versus subsequent Population II stars, with results for a specific model parameter value.}{2}
\qrow{notation}{[Fe/H] or [Fe] from $m_{\rm PISN}$ and Pop II progenitors}{2}
\end{querytable}
\end{fotoexample}

\begin{fotoexample}{2304.10129}{15}
\extitle{Early-time spectroscopic modelling of the transitional Type Ia Supernova 2021rhu with TARDIS}
\excaption{The result of the energy scaling upon the N100 density profile. The profiles with lower kinetic energy are less able to accelerate the ejected material to high velocities and therefore possess a more compact ejecta structure. The black dotted line represents the base N100 density profile with the 30000 extension. The inset displays the location of the high velocity tail of the preferred 60 per cent scaled density profile between the W7 and double-detonation profiles.}
\begin{querytable}
\qrow{terse}{kinetic energy scaling density profile velocity comparison}{18}
\qrow{casual}{How does changing the kinetic energy affect the density profile and velocities in the ejecta?}{N/A}
\qrow{vague}{effect of energy on density structure and ejected material velocities}{N/A}
\qrow{detailed}{Comparison of density profiles resulting from different kinetic energy scalings, showing how lower energy cases produce slower ejecta velocities and different density structures.}{8}
\qrow{notation}{$\rho(v)$ profiles: $E_{\rm kin}$ scaling effects on $v_{\rm eject}$ and density acceleration}{10}
\end{querytable}
\end{fotoexample}

\begin{fotoexample}{0705.4195}{4}
\extitle{VSOP: The Variable Star One-shot Project I. Project presentation and first data release}
\excaption{Selected spectral regions of V1045 Sco: Top: H$\alpha$ profile. Note the emission on both the red and blue edges, and the strong reversal. Bottom: The lithium line region.}
\begin{querytable}
\qrow{terse}{V1045 Sco H-alpha lithium spectral profiles}{1}
\qrow{casual}{Can you find that figure showing the H-alpha and lithium lines for V1045 Sco with that weird double-peaked emission?}{1}
\qrow{vague}{spectral regions showing emission line with reversal and lithium absorption feature}{95}
\qrow{detailed}{Two spectral regions of V1045 Sco: the top panel displays the H-alpha line showing emission on both red and blue edges with strong central reversal; the bottom panel shows the lithium line region.}{1}
\qrow{notation}{V1045 Sco H$\alpha$ profile: asymmetric emission with reversal; Li line region}{1}
\end{querytable}
\end{fotoexample}

\begin{fotoexample}{1706.05004}{1}
\extitle{KiDS+GAMA: Cosmology constraints from a joint analysis of cosmic shear, galaxy-galaxy lensing and angular clustering}
\excaption{Normalised redshift distribution of the four tomographic source bins of KiDS (solid lines), used to measure the weak gravitational lensing signal, and the normalised redshift distribution of the two spectroscopic samples of GAMA galaxies (histograms), that serve as the foreground sample in the galaxy-galaxy lensing analysis and that are used to determine the angular correlation function. For plotting purposes, the redshift distribution of GAMA galaxies has been multiplied by a factor 0.5. The shaded regions indicate the photometric redshift ($z_{\rm B}$) selection of the tomographic source bins.}
\begin{querytable}
\qrow{terse}{Source lens redshift distributions tomographic bins}{N/A}
\qrow{casual}{Can you find the plot showing the redshift distributions for the different source and lens samples?}{N/A}
\qrow{vague}{Redshift distributions of galaxy samples used in weak lensing analysis}{N/A}
\qrow{detailed}{Plot displays normalized redshift distributions for multiple tomographic source bins shown as solid lines, compared against lens sample redshift distributions across the range of observed redshifts.}{N/A}
\qrow{notation}{$n(z)$ for source bins versus lens bins, normalized distributions}{N/A}
\end{querytable}
\end{fotoexample}

\begin{fotoexample}{astro-ph/0702176}{6}
\extitle{Nucleosynthesis and Remnants in Massive Stars of Solar Metallicity}
\excaption{Elemental yields integrated over a Salpeter initial mass function for solar metallicity stars with masses from 12 to 120 . Calculations were carried out for for two choices of explosion energy (1.2, 2.4 B) and two piston location (at the place where the entropy jumps to $S/N_Ak = 4$ and at the edge of the iron core). Only minor differences are discernible. Similarly small changes occur when the slope of the IMF is changed from $-1.35$ (Salpeter) to a gentler $-0.9$.}
\begin{querytable}
\qrow{terse}{elemental yields Salpeter IMF solar metallicity massive stars}{1}
\qrow{casual}{Can you find that plot showing how much of each element gets produced by massive stars at solar metallicity?}{5}
\qrow{vague}{integrated yields across initial mass function for different explosion energies in massive stars}{3}
\qrow{detailed}{Plot displaying elemental yields integrated over a Salpeter initial mass function for solar metallicity stars between 12 and 120 solar masses, comparing results for two different explosion energy choices.}{1}
\qrow{notation}{$Y_i(Z = Z_\odot)$ vs $A$ for Salpeter IMF with varied $E_{\rm SNe}$}{N/A}
\end{querytable}
\end{fotoexample}

\begin{fotoexample}{1305.0309}{11}
\extitle{The Influence of Motion and Stress on Optical Fibers}
\excaption{The 4 position parameters (X, Y, rho, lower stage) plotted against one another in order to better understand the correlations between the positions of the test platform and the transient FRD seen in the lifetime tests. The SOE value is indicated by color, ranging from zero (plotted as black) to the highest SOE values (plotted as light green to red). There are two primary sections along the path (both turnaround points for the Y motion) where FRD events are seen to increase. Figures \texttt{<ref>}, \texttt{<ref>} and \texttt{<ref>} plot the results for 3 different tracks and rates in the same manner as shown here.}
\begin{querytable}
\qrow{terse}{Position parameters correlation matrix optical fiber}{27}
\qrow{casual}{Can you find that figure showing how all the position parameters relate to each other in the fiber test setup?}{1}
\qrow{vague}{Plot showing correlations between different position measurements in the experimental apparatus}{N/A}
\qrow{detailed}{A figure displaying pairwise correlations among four position parameters (X, Y, rho, lower stage) plotted against one another to reveal relationships between test platform positions and transient optical effects.}{8}
\qrow{notation}{Correlation analysis of \{X, Y, $\rho$, lower stage\} parameters vs.\ transient FRD response}{1}
\end{querytable}
\end{fotoexample}

\begin{fotoexample}{2201.09905}{7}
\extitle{The Effect of Stellar Contamination on Low-resolution Transmission Spectroscopy: Needs Identified by NASA's Exoplanet Exploration Program Study Analysis Group 21}
\excaption{Maps of the vertically directed bolometric intensity for 12 of the 18 magnetic simulation runs. Note that the scale of the normalized intensity is adjusted for each spectral type for improved image contrast, and the grayscales saturate at the limits indicated for each row. As the average magnetic field strength increases, more MBPs are visible within the inter-granular lanes; the quantity of MBPs also increases with effective temperature, but decreases again at temperatures greater than solar. From \texttt{<cit.>}.}
\begin{querytable}
\qrow{terse}{Bolometric intensity maps magnetic simulation runs spectral types}{1}
\qrow{casual}{Can you find those intensity maps from the magnetic simulations across different stellar types?}{66}
\qrow{vague}{Maps showing intensity distribution for multiple stellar simulations with different normalized scales}{N/A}
\qrow{detailed}{Figure displays bolometric intensity maps from magnetic simulations across various spectral types, with normalized intensity scales adjusted individually for each type to enhance visibility of the spatial patterns.}{1}
\qrow{notation}{$I({\rm vertical})$ maps for $B_{\rm field}$ simulations, normalized per spectral class}{N/A}
\end{querytable}
\end{fotoexample}

\begin{fotoexample}{2110.03638}{4}
\extitle{Massive Star Cluster Formation and Destruction in Luminous Infrared Galaxies in GOALS II: An ACS/WFC3 Survey of Nearby LIRGs}
\excaption{The age distribution function for NGC 7469 which contains 146 confirmed SSCs. After excluding clusters in the nuclear ring, we find that the distribution is consistent with the overall slope measured for the full sample, and importantly, is found to be steeper than what is measured for lower-luminosity star-forming galaxies like the LMC.}
\begin{querytable}
\qrow{terse}{NGC 7469 age distribution function star clusters}{1}
\qrow{casual}{Do you have that plot showing the ages of star clusters in NGC 7469?}{1}
\qrow{vague}{Age distribution histogram of star clusters excluding nuclear region}{N/A}
\qrow{detailed}{A histogram showing the age distribution function of star clusters in NGC 7469, with 146 confirmed clusters plotted, excluding those in the nuclear ring to demonstrate the overall slope of the distribution.}{1}
\qrow{notation}{$dN/d(\log {\rm age})$ for SSCs in NGC 7469, excluding nuclear ring}{1}
\end{querytable}
\end{fotoexample}

\begin{fotoexample}{2202.02773}{8}
\extitle{Probing Cosmic Inflation with the LiteBIRD Cosmic Microwave Background Polarization Survey}
\excaption{Payload module overview showing the surrounding V-grooves acting as passive coolers up to 30 K, with the 4.8-K enclosure of the low-frequency crossed-Dragone telescope in the front, and the two 4.8-K tubes of the mid- and high-frequency on-axis telescopes in the back.}
\begin{querytable}
\qrow{terse}{LiteBIRD payload module cryogenic architecture diagram}{10}
\qrow{casual}{Can you show me the diagram of how the LiteBIRD instrument is laid out with all its cooling components?}{18}
\qrow{vague}{Schematic showing telescope and cooler components at different temperature stages}{N/A}
\qrow{detailed}{Technical diagram illustrating the payload module architecture with V-groove passive coolers operating at 30 K and low-frequency telescope enclosure at 4.8 K, showing spatial arrangement of detector systems.}{1}
\qrow{notation}{$T \sim 4.8$ K, $T \sim 30$ K cryogenic payload configuration diagram}{86}
\end{querytable}
\end{fotoexample}

\begin{fotoexample}{2206.04999}{2}
\extitle{Environmental variation of the low-mass IMF}
\excaption{Number of stars formed as a function of star formation efficiency. The solid lines are realisation 1 and the dashed lines are realisation 2.}
\begin{querytable}
\qrow{terse}{Star formation efficiency versus number of stars formed}{7}
\qrow{casual}{How does the number of stars change with different star formation efficiencies across two different simulation runs?}{2}
\qrow{vague}{Plot showing how stellar population varies with formation efficiency in two separate simulations}{28}
\qrow{detailed}{A graph with star formation efficiency on one axis and number of stars on the other, comparing two realizations shown as solid and dashed lines respectively.}{1}
\qrow{notation}{$N_*$ vs $\epsilon_{\rm ff}$: two realizations plotted as continuous and dashed curves}{N/A}
\end{querytable}
\end{fotoexample}

\begin{fotoexample}{2102.05601}{3}
\extitle{No Evidence for Orbital Clustering in the Extreme Trans-Neptunian Objects}
\excaption{The black contours (which follow a linear scale) are Gaussian kernel density estimates of $10^6$ iterations of sampling 5 points from the OSSOS biases calculated using the CFEPS/OSSOS survey simulator (because OSSOS discovered 5 ETNOs), and plotting the mean $(x,y)$ and $(p, q)$ positions (see 4 for definitions of $x$, $y$, $p$, and $q$). The red contours represent the same statistic, calculated using . The blue points are the mean positions of the 5 ETNOs discovered by OSSOS. Our simulator reproduces the results of the OSSOS simulator with better fidelity than the heuristic method of BB19 (see their Figure 4).}
\begin{querytable}
\qrow{terse}{kernel density estimates OSSOS survey simulator sampling}{1}
\qrow{casual}{Can you find that figure showing the density contours from the OSSOS survey simulator with the sampled points?}{1}
\qrow{vague}{contour plot showing kernel density estimates from survey simulator bias calculations}{13}
\qrow{detailed}{Black contours representing Gaussian kernel density estimates plotted on a linear scale, generated from one million iterations of sampling five points from survey simulator biases. The figure visualizes the distribution results from the CFEPS/OSSOS survey simulation.}{1}
\qrow{notation}{KDE contours from $10^6$ Monte Carlo samples of 5-point selections with CFEFS/OSSOS simulator bias}{3}
\end{querytable}
\end{fotoexample}

\begin{fotoexample}{1904.01581}{6}
\extitle{The HI mass function in the Parkes HI Zone of Avoidance survey}
\excaption{Schechter parameters for the four sub-samples divided according to their local density projected on the $\alpha-\log M_{HI}^*$ plane. Ellipses are defined with $(\alpha,\log M_{HI}^*,\sigma_{\alpha},\sigma_{\log M_{HI}^*},\theta)$ and plotted on a colour scale from the least dense regions (blue) to the most dense regions (red). Dark and light ellipses show $1\sigma$ and $3\sigma$ contours, respectively.}
\begin{querytable}
\qrow{terse}{Schechter parameters alpha versus HI mass density bins}{3}
\qrow{casual}{Can you find that plot showing how the Schechter parameters change across different density environments?}{N/A}
\qrow{vague}{Schechter function parameters plotted for different density samples with ellipse uncertainties}{2}
\qrow{detailed}{Two-dimensional plot with alpha on the y-axis and log HI mass on the x-axis, showing Schechter parameters for four density-divided sub-samples marked by ellipses representing parameter uncertainties.}{1}
\qrow{notation}{$\alpha$ vs $\log M_{\rm HI}^*$ plane with $\sigma_\alpha$ confidence ellipses for local density sub-samples}{1}
\end{querytable}
\end{fotoexample}

\begin{fotoexample}{astro-ph/0508301}{6}
\extitle{The Arecibo Legacy Fast ALFA Survey: I. Science Goals, Survey Design and Strategy}
\excaption{Sky distribution of expected detections for the ALFALFA survey, as rendered with a Z97 HIMF. In panel (a) sources of all HI masses are plotted, while in panel (b) only those with $M_{HI} < 10^8$ $M_\odot$ are shown. Note that the ALFALFA survey will be restricted to Right Ascensions $07^h$ to $16.5^h$ and $22^h$ to $3^h$, although the full range of R.A.s is shown in the figure.}
\begin{querytable}
\qrow{terse}{ALFALFA survey sky distribution expected detections HIMF}{1}
\qrow{casual}{Can you find the figure showing where we expect to detect sources across the sky in ALFALFA?}{10}
\qrow{vague}{Sky map showing distribution of expected HI detections split by mass limits}{N/A}
\qrow{detailed}{Two-panel figure displaying sky distribution of expected detections using a mass function model, with panel (a) showing all HI masses and panel (b) showing only lower mass sources below a threshold.}{N/A}
\qrow{notation}{Sky distribution $N(M_{\rm HI})$ from HIMF $z \sim 0$, $M_{\rm HI} < 10^8\,M_\odot$ comparison}{N/A}
\end{querytable}
\end{fotoexample}
\end{document}